\documentstyle[11pt,newpasp,twoside,epsf]{article}
\begin{document}
\title{Kinematics of Extremely Faint Dwarf Galaxies}
 \author{Ayesha Begum}
\affil{National Centre for Radio Astrophysics,
        Post Bag 3, Ganeshkhind, Pune 411 007}
\author{Jayaram N. Chengalur}
\affil{National Centre for Radio Astrophysics,
        Post Bag 3, Ganeshkhind, Pune 411 007}

\begin{abstract}

We present the results of deep, high velocity resolution ($\sim$1.6 kms$^{-1}$) 
Giant Meterwave Radio Telescope (GMRT)  HI 
21cm observations of extremely faint (M$_{\rm{B}} > -$12.5) dwarf irregular galaxies.
We find that all of our sample galaxies show systematic large scale velocity gradients,
unlike earlier studies which found chaotic velocity fields for such faint galaxies. For some 
of the sample galaxies the velocity fields are completely consistent with ordered rotation, 
though the peak circular velocities are comparable to the observed  random motions. These are
the faintest known galaxies with such regular kinematics. We present (``asymmetric drift" corrected) 
rotation curves  and mass models (including fits for Isothermal and NFW halos) for some of these 
galaxies and discuss the implications for hierarchical models of galaxy formation.

\end{abstract}

\section{Introduction}
\label{sec:intro}

Dwarf irregular galaxies are typically dark matter dominated throughout, 
unlike brighter spirals, where both stars and gas make a significant contribution 
to the dynamical mass in the inner regions.
Sensitive studies of the kinematics of the faintest dwarf systems 
thus provide direct information on the density profiles of their dark matter halos and can
 hence be used to place constraints on models of structure formation.
However, a major stumbling block in such programs is that it is still unclear 
whether very faint dwarf irregular galaxies are rotationally supported or not.
From a systematic study of the kinematics of a sample of dwarfs, C\^{o}t\'{e}, Carignan 
\& Freeman  (2000)  found that normal rotation is  seen only in galaxies 
brighter than -14 mag, while fainter dwarfs have disturbed kinematics. 
This result is consistent with the earlier findings of Lo, Sargent \& Young (1993), who 
from a study of a sample of dwarfs  (with M$_{B\rm} \sim -9$ to M$_{B\rm} \sim -15$)
found that very faint dwarf irregulars have chaotic velocity fields.
However, most of these earlier studies were done with  modest velocity 
resolutions ($\sim 6-7$ kms$^{-1}$) and modest sensitivities which makes 
it difficult to discern systematic patterns (which typically have amplitudes 
$<$ 10 kms$^{-1}$), if any, in the velocity fields of such faint systems.  
We present here high velocity resolution ($\sim$1.65~kms$^{-1}$), GMRT HI 21
cm observations of a sample of extremely faint dwarf galaxies.

\section{ The GMRT Sample }
\label{sec:sample}

Our GMRT sample consist of dwarf irregular galaxies with 
M$_{\rm{B}}> -12.5$ mag and have typical HI masses of 
$\sim10^{6-7}$M$_\odot$. 
The typical integration time on each source is~$\sim16-18$ hrs, 
which gives a typical rms of~$\sim$1.0 Jy/Beam per channel.

\section{Results}
\label{sec:res}

       Unlike earlier studies, our high velocity resolution 
and high sensitivity GMRT observations detect  large
scale systematic patterns in the velocity fields of our
sample galaxies (see e.g. Fig.\ref{fig:overlay},\ref{fig:mom1}). 
For some of the galaxies, the large scale
gradients can be modelled as systematic rotation, allowing
us to derive the rotation curves and hence determine
the structure of their dark matter halos from mass modeling.
Rest of this paper  discusses the results obtained from the detailed
kinematical study of two of our sample galaxies, Camelopardalis B
(Cam B) and DDO210.

\subsection{ Camelopardalis B }
\label{sssec:camb}

\begin{figure}
\plotfiddle{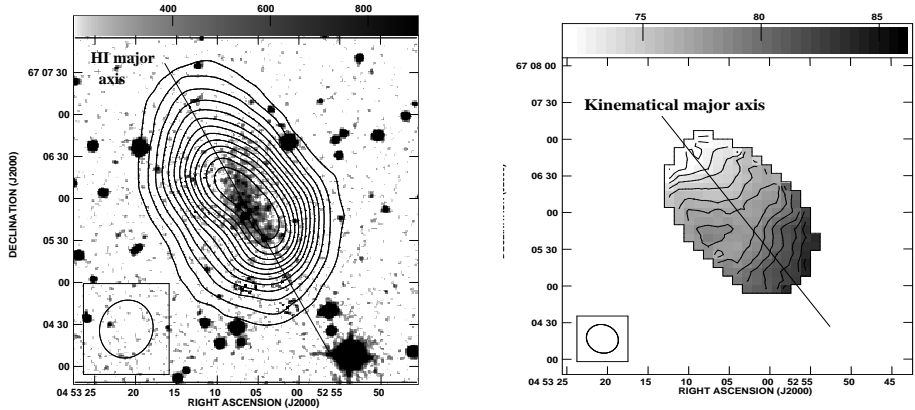}{2.0in}{-90}{70}{70}{-290}{290}
\caption{{\bf{(Left)}}Integrated HI emission map of Cam B at
         $40^{''} \times 38^{''}$ resolution 
          overlayed on the Digitised Sky Survey Image.
          The contour levels are 0.05, 0.12,0.19,0.26,0.33
          0.41,0.50,0.55,0.62,0.69,0.76,0.83,0.89 Jy/Beam kms$^{-1}$.
          {\bf{(Right)}}The HI velocity field of Cam~B at $24^{''}\times 22^{''}$ arcsec
          resolution. The contours are in steps of 1~kms$^{-1}$ and
          range from 70.0~kms$^{-1}$ (the extreme North West contour)
          to 84.0~kms$^{-1}$ (the extreme South East contour).
         Note that the kinematical major axis of the galaxy is well
  aligned with the HI and optical major axis.}

\label{fig:overlay}
\end{figure}

\begin{figure}
\plotfiddle{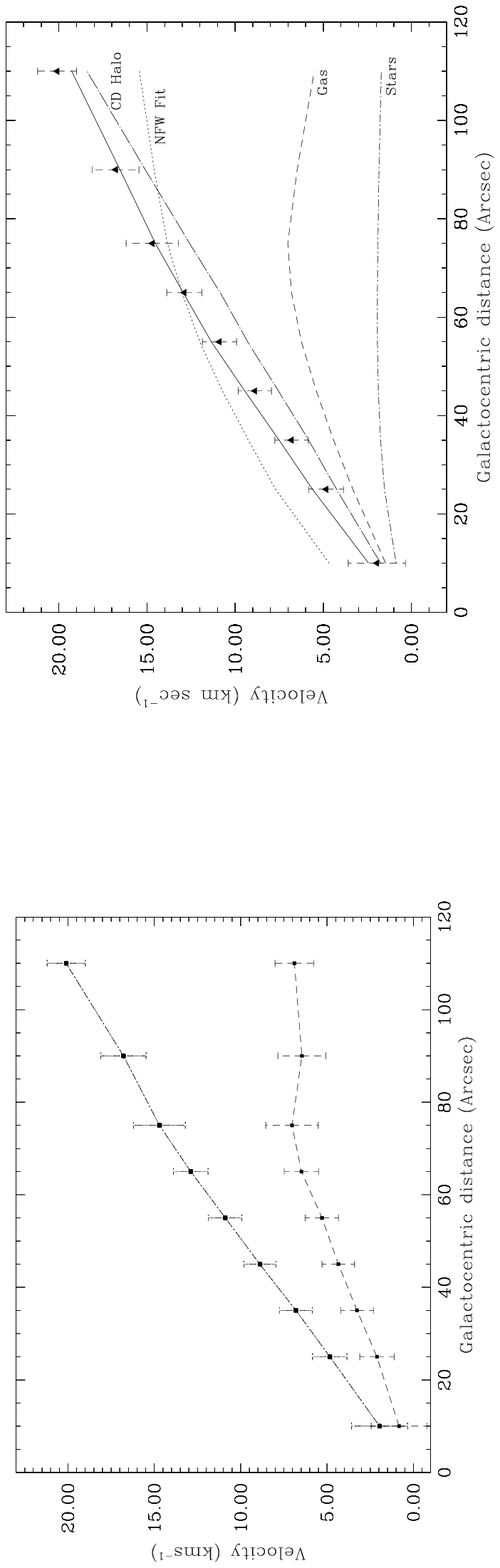}{1.8in}{-90}{60}{70}{-250}{280}
\caption{{\bf{(Left)}}The derived rotation curve for Cam B (dashes)
and the rotation curve after applying
          the asymmetric drift correction (dash dots).
{\bf{(Right)}}Mass models for Cam B using the corrected rotation curve.
The points are the observed data. The total mass of gaseous disk (dashed line)
is $6.6\times10^6 M_\odot$.The stellar disk (short dash dot line) has
$\Upsilon_V=0.2$, giving a stellar mass of $0.7 \times10^6 M_\odot$. The
best fit total rotation curve for the constant density halo model is shown as
a solid line, while the contribution of the halo itself is shown as a
long dash dot line (the halo density is $\rho_0=13.7\times10^{-3}
M_\odot$ pc$^{-3}$). The best fit total rotation curve for an NFW type halo
(for $c=1.0$ and $\Upsilon_V=0.0$) is shown as a dotted line. 
}
\label{fig:corr_curve}
\end{figure}

Cam B is a very  faint (M$_B \sim -10.9$) dwarf irregular
galaxy at a distance of 2.2 Mpc. Fig.~\ref{fig:overlay} shows 
the integrated HI emission from the galaxy at $40^{''}\times38^{''}
$~arcsec resolution overlayed on the digitized sky survey image. 
The HI emission extends out to a galacto-centric distance $>$ 4 
times the optical scale length. We have estimated an HI mass 
of 5.3$\pm0.5 \times{10}^{6} M_\odot$ (taking the  
distance to the galaxy to be 2.2 Mpc) from the integrated global 
HI emission profile of the galaxy.

       The velocity field of Cam B derived from the moment analysis of
$24^{''}\times22^{''}$ resolution data is shown in Fig.~\ref{fig:overlay}. 
As can be seen that inspite of being very faint, Cam B shows a very regular kinematics.
The  isovelocity contours are approximately parallel, which is a signature of 
rigid body rotation. Also, the  kinematic major axis of the galaxy is well 
aligned with the major axis of both the HI distribution and the optical 
image. Cam~B is the faintest known dwarf irregular galaxy to show such  regular 
kinematics. 
     
        The rotation curve for Cam B was derived from the HI velocity field, 
assuming it to be an axi-symmetric disk (details can be found in Begum, 
Chengalur \& Hopp 2003). The dashed curve in Fig.~\ref{fig:corr_curve}  
shows the derived rotation curve. The peak (inclination 
corrected) rotation velocity for Cam B is $\sim$ 7 kms$^{-1}$,
comparable to the observed dispersion of $\sim$ 7 kms$^{-1}$ in the HI gas. This
implies that random motions provide significant dynamical support to Cam B.
Equivalently, the observed circular velocity underestimates the dynamical mass
of the galaxy. Hence, before constructing mass models, the observed rotational 
velocity was corrected for the pressure support using the usual ``asymmetric drift
 correction" (see  Begum et al.(2003) for more details) . The dot-dashed curve in 
Fig.~\ref{fig:corr_curve} shows the ``asymmetric drift" corrected rotation curve. 

Using this corrected curve, mass models for Cam B were derived.  
Fig.~\ref{fig:corr_curve}  shows the best fit mass model 
using a modified isothermal halo. Also shown in the figure is 
the best fit total rotation curve for an NFW type halo. 
As can be seen, the kinematics of Cam~B is  well fit with 
a modified isothermal halo while an NFW halo provides 
a poor fit to the data.

 The ``asymmetric drift" corrected rotation curve for Cam B is 
rising till the last measured point (Fig.~\ref{fig:corr_curve}), hence the core 
radius of the isothermal halo could not be constrained 
from the data. The best fit model for a constant density halo 
gives central halo density ($\rho_0$) of 12.0$\times 10^{-3}~M_\odot~pc^{-3}$. 
The derived $\rho_0$ is relatively insensitive to the assumed
mass-to-light ratio of the stellar disk ($\Upsilon_V$). We found that 
by changing  $\Upsilon_V$ from a value of 0 (minimum disk fit) 
to a value of 2.0 (maximum disk fit), $\rho_0$ changes by $<$20\%, 
hence is well determined.  From the last measured
point of the  observed rotation curve, a total dynamical mass of
1.1$\times 10^8 M_\odot$ is derived, i.e. at the last measured point
more than 90\% of the mass  of Cam~B is dark. Futher, the dominance of the 
dark matter halo together with the linear shape of the rotation 
curve (after correction for ``asymmetric drift'') means that one 
cannot obtain a good fit to the rotation curve using an NFW halo 
regardless of the assumed $\Upsilon_V$.

\subsection{ DDO210 }
\label{ssec:ddo210}

\begin{figure}
\plotfiddle{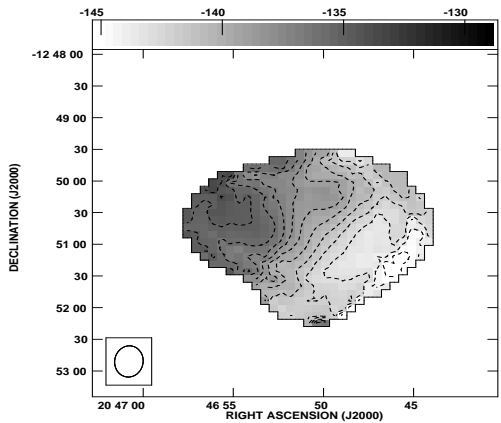}{2.2in}{-0}{35}{30}{-130}{-40}
\caption{The HI velocity field of DDO210 at 29$^{''}\times 23^{''}$
          resolution. The contours are in steps of 1~kms$^{-1}$ and
          range from $-$145.0~kms$^{-1}$ to $-$133.0~kms$^{-1}$.
        }
\label{fig:mom1}
\end{figure}

\begin{figure}
\plotfiddle{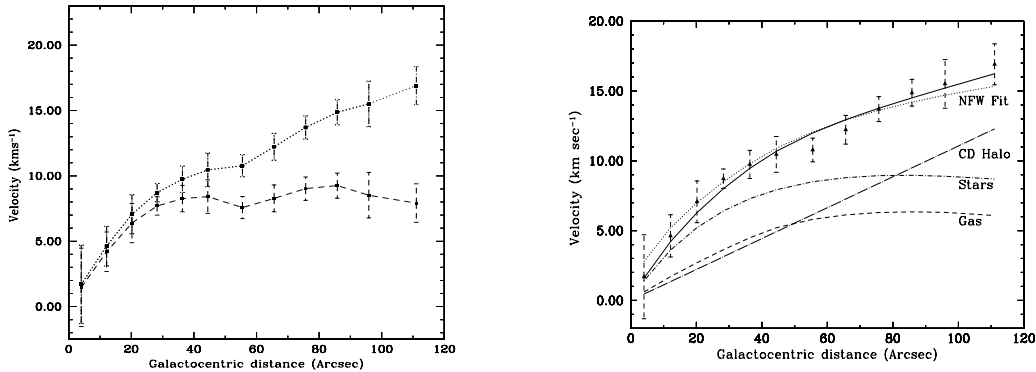}{1.8in}{-90}{68}{70}{-270}{285}
\caption{{\bf{(Left)}}The hybrid rotation curve (dashes)  and the rotation curve
        after applying the asymmetric drift correction (dots).
{\bf{(Right)}} Mass models for DDO210 using the corrected rotation curve.
         The points are the observed data. The total mass of gaseous
            disk (dashed line)  is $3.6\times10^6 M_\odot$.The stellar 
          disk (short dash dot line) has
          $\Upsilon_B=3.4$, giving a stellar mass of $9.2 \times10^6 M_\odot$. 
        The  best fit total rotation curve for the constant density 
          halo model is shown as
          a solid line, while the contribution of the halo itself 
           is shown as a  long dash dot line (the halo density is 
           $\rho_0=29~\times10^{-3} M_\odot$ pc$^{-3}$).
         The best fit total rotation curve for an NFW type halo,
          using $\Upsilon_B=0.5$, c=5.0 and $v_{200}$=38.0~km s$^{-1}$
    is shown as a dotted line. See text for more details.
  }
\label{fig:v_asy}
\end{figure}

DDO210 is the faintest (M$_B \sim -10.6$) known dwarf irregular galaxy 
in the local group. From a recent HST observations  Karachentsev et al. (2002) 
have estimated a distance to the galaxy to be 950$\pm$50 kpc. The HI mass of 
DDO210, determined from the global HI emission profile from our 
observations  is $2.8\pm0.3  \times{10}^{6} M_\odot$ (taking 
the distance to the galaxy to be 1 Mpc).

Fig.~\ref{fig:mom1} shows the velocity field of DDO210 derived from  
29$''\times 23''$ resolution data cube. The velocity field is regular  
and a systematic velocity gradient is seen across the galaxy.
This galaxy was also  a part of the sample of Lo et al.(1993).
However, our velocity field differs significantly from the velocity field 
derived by  Lo et al.(1993). The systematic pattern seen in our velocity 
field is, to zeroth order, consistent with that expected from 
rotation. On the other hand, the velocity field derived by Lo et al.(1993) 
(based on a coarser  velocity resolution of $\sim$ 6 kms$^{-1}$) is chaotic.  
This difference in the observed kinematics suggests that high velocity resolution 
and high sensitivity is  crucial in determining the systematic gradients in 
the velocity fields of faint galaxies like DDO210.

 Fig.~\ref{fig:v_asy} shows the derived rotation curve for DDO210 and the 
``asymmetric drift" corrected rotation curve.  We find that the ``asymmetric
drift" corrected rotation curve of DDO210  can be well fit with  either a 
modified isothermal halo (with a central density $\rho_0 \sim 29\times10^
{-3}$ $M_\odot$ pc$^{-3}$) or an NFW halo. In the case of the NFW halo however, 
a good fit is obtained for a wide range of  parameters; the halo parameters 
could not be uniquely determined from the fit. Fig.~\ref{fig:v_asy} shows 
the best fit mass model for DDO210. 

\section{ Discussion }
\label{sec:discuss}

\begin{figure}[]
\plotfiddle{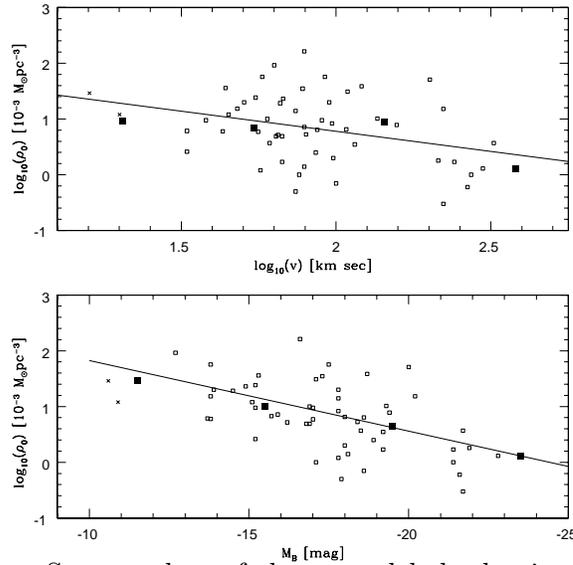}{2.7in}{0}{40}{40}{-135}{-80}
\caption{Scatter plots of the central halo density against the
          circular velocity and the absolute blue magnitude . The data (empty squares)
          are from Verheijen~(1997), Broeils~(1992),
          C\^{o}t\'{e} et al.~(2000) and Swaters~(1999). The filled
          squares are the medians of the binned data, and the straight
          lines are the best fits to the data. Cam~B and DDO210 are
            shown  as crosses.
        }
\label{fig:dens}
\end{figure}

In Fig.~\ref{fig:dens} we plot the core density of isothermal halo against circular
velocity and absolute blue magnitude for a sample of galaxies, spanning a range of
magnitudes from M$_B\sim10.0$ mag to M$_B\sim23.0$ mag. Cam B and DDO210 are also
shown in the figure, lying at the low luminosity end of the sample. We have 
used B magnitudes because this is the only band for which data is currently available 
for both  our sample and the other galaxies. As can be seen in the figure, 
there is a trend of increasing halo density with a decrease in circular velocity 
and absolute magnitude, shown by a best fit line to the data (solid line), 
although the correlation is very weak and noisy. Further, as a guide to an 
eye, we have also binned the data and plotted the median value (solid points). 
Binned data also shows a similar trend. Such a tread is expected in 
hierarchical structure formations scenario (e.g. Navarro, 
Frenk \& White 1997).

\end{document}